\documentstyle [12pt,epsf] {article} 
    
   \catcode`@=12 
\begin{document} 
   \vspace{0.5in} 
   \oddsidemargin -.375in 
   \newcount\sectionnumber 
   \sectionnumber=0

\def\bea{\begin{eqnarray}}
\def\eea{\end{eqnarray}}
\def\slash#1{\not  \! \! {#1}}
\def\mgut{M_{\rm GUT}}
\def\agut{\alpha_{\rm GUT}}
\def\gtilu{{\tilde g}_u}
\def\gtild{{\tilde g}_d}
\def\gtilup{{\tilde g}_u^\prime}
\def\gtildp{{\tilde g}_d^\prime}
\def\gtiluq{{\tilde g}_u^2}
\def\gtildq{{\tilde g}_d^2}
\def\gtilupq{{\tilde g}_u^{\prime 2}}
\def\gtildpq{{\tilde g}_d^{\prime 2}}
\def\gtiluc{{\tilde g}_u^3}
\def\gtildc{{\tilde g}_d^3}
\def\gtilupc{{\tilde g}_u^{\prime 3}}
\def\gtildpc{{\tilde g}_d^{\prime 3}}
\def\gtiluqq{{\tilde g}_u^4}
\def\gtildqq{{\tilde g}_d^4}

   \def\be{\begin{equation}} 
   \def\ee{\end{equation}} 
   \def\ber{\begin{eqnarray}} 
   \def\eer{\end{eqnarray}} 
\def\vn{\vec{n}} 
   \thispagestyle{empty} 
   \begin{flushright} 
%UT-PT-99-18\\ October 1999\\
   \end{flushright} 
   \vspace {.5in} 
   \begin{center} 
   {\Large \bf Vacuum Stability in Split Susy and Little Higgs Models \\} 
\vspace{.5in} 
   {\rm Alakabha Datta$^1$ and Xinmin Zhang$^2$ \\} 
   \vspace{.5in} 
   {$^1$ \it Department of Physics, University of Toronto, \\ 
   Toronto, Ontario, Canada, M56 1A7\\} 

   {$^2$ \it CCAST (World Laboratory), P.O. Box 8730, Beijing 100080 and\\ 
   Institute of High Energy Physics, Beijing 100039, P.R. China\\} 
   \vspace{.1in} 
   \vskip .5in 
   \end{center} 
   \begin{abstract}

We study the stability of the effective Higgs potential in the split
supersymmetry  and  Little Higgs models.
In particular, we study the effects of higher dimensional operators
in the effective potential on the Higgs mass predictions. We find that the size 
and sign of the higher dimensional operators can significantly change the
Higgs mass required to maintain vacuum stability in Split Susy models.
In the Little Higgs models the effects of higher dimensional operators
can be large because of a relatively lower cut-off scale. Working with a specific model
we find that a  contribution from the 
higher dimensional operator with coefficient of $O(1)$ can destabilize the vacuum.
\end {abstract} 
\newpage 
\baselineskip 24pt 

\section{\bf Introduction} 

    In the standard model (SM),
   the 
   electroweak symmetry breaking is achieved through  a complex 
   fundamental Higgs scalar. The discovery of the Higgs is
    one of the chief goals of the upcoming LHC experiment.
   Arguments 
   of "triviality" \cite{1} and "naturalness" \cite {2}, 
   suggest that the simple spontaneous 
   symmetry breaking mechanism in the SM  may 
   not be the whole story. 
   In the SM, the Higgs mass receives quadratically divergent quantum corrections
   which has to be canceled by some new physics(NP) to obtain a sensible Higgs mass.
    The Higgs sector of the 
   standard model is therefore an effective theory valid below 
   some cut-off scale $\Lambda$. 
   For the cancellation of the mass divergence to be not finely tuned one would require the NP
   scale $ \Lambda$ to be $ \sim $ TeV. One of the popular models of NP is 
   supersymmetry(SUSY) and the phenomenology of  low  energy SUSY- that is SUSY broken at
   a scale $ \sim $ TeV, has been
   an active area of research for a long time now. However, there is no evidence of SUSY yet. As
    various experiments increasingly constrain the parameter space of the minimal models
    of low energy SUSY such as MSSM it is possible that SUSY, if present in nature, 
    may manifest itself in experiments in a manner very different from what has been expected so far.
    An interesting model of SUSY with very different phenomenology than the usually studied SUSY models was recently 
    proposed in Ref. \cite{savas1,savas2}, which was coined the name split supersymmetry in Ref\cite{Guidice}.
In this scenario, the Higgs mass is finely tuned to be at the weak scale while all the other scalars 
are much heavier than the electroweak scale. The fermions in the SUSY spectrum are allowed to be 
light-around a TeV or less due to chiral symmetries.  This scenario can provide a dark matter candidate and
 the possibility of observing SUSY particles at LHC remains. By making the scalars heavy FCNC effects
 at low energies are suppressed. However, there are some hints of new FCNC effects 
 in B-decays\cite{Datta1,Datta2,Datta3,Datta4,Datta5}
 and if they are confirmed then Split Susy models will have to be modified to accommodate them.
 Nonetheless, Split Susy models have some novel features and interesting phenomenology that have been 
 already explored 
 in the 
 literature\cite{sspheno1,sspheno2,sspheno3,sspheno4,sspheno5,sspheno6,sspheno7,sspheno8,sspheno9,sspheno10,
 sspheno11,sspheno12,sspheno13,sspheno14,sspheno15,sspheno16,sspheno17,sspheno18,vacsusy}.

There are also non supersymmetric NP that can cure the Higgs mass problem. 
In the Little Higgs models \cite{ACG2,Arkani-Hamed:2002qx,
Arkani-Hamed:2002qy,Gregoire:2002ra,Low:2002ws,Kaplan:2003uc,Schmaltz:2002wx,Chang:2003un,
Skiba:2003yf,Chang:2003zn,Cheng:2003ju,Cheng:2004yc,Martin} the Higgs mass divergence is canceled by contributions 
from additional particles with masses around a TeV. In this scenario, the Higgs is a 
pseudo Nambu-Goldstone boson
whose mass is protected by some global symmetry which is spontaneously broken. The 
Higgs acquires a mass from explicit breaking of the global symmetry.
   
Whatever model one considers as a candidate for beyond the SM physics, the predictions of the Higgs mass
is a key prediction of that candidate model.
   Direct searches  at LEP  has put a
lower bound on the Higgs mass $m_h \sim 115$ GeV \cite{PDG,LEP} and 
from a global fit to
electroweak precision 
   observables one can put an upper bound on the Higgs mass, $m_h <$ 246 GeV
   at 95 \% C.L \cite{PDG,LEP}. 
   This assumes that the scale of new physics, $ \Lambda$, 
   is high enough and thus new physics does not have 
significant effects on the electroweak 
   precision observables. 
This upper bound on the Higgs mass 
   may be relaxed if 
   the scale $\Lambda$, which suppresses the higher 
   dimensional operators arising from new physics, is around a few 
TeV \cite {Hall,Barb,Ravi}. There are also
theoretical arguments of triviality and vacuum stability which
place bounds on the Higgs mass.
 A lower bound on the Higgs mass $\sim 135$ GeV  is
   obtained 
   by requiring the standard model vacuum to be stable to the Planck
scale \cite{Casas1,Casas2,Casas3,Casas4}.
  It was shown in Ref
   \cite{DYZ,DZ,others1, others2} 
   that, in the presence of higher dimensional operators, the vacuum
stability limit on the Higgs mass can also be changed.
In this work we will study the effects of higher dimensional operators, due to physics
around the cut-off, on the Higgs mass prediction in Split Susy and 
the Little Higgs
 model. It is worthwhile to justify the addition of higher dimensional operator
for vacuum stability analysis. 
Consider the Split SUSY scenario where the
particle spectrum is split with the scalars having masses close to the 
SUSY breaking scale, $m_S$. These scalars as well as other possible 
SUSY multiplet at or around $m_S$ will contribute to the Higgs 
effective potential and at low energy such effects can be represented by an
dimension 6 effective
operator. What we point out in the paper is that the effect of this higher 
dimension operator can be significant for Higgs mass prediction.
The same arguments apply to the Little Higgs model where additional fermions 
have to be placed around the cut-off to cancel anomalies and 
for phenomenological reasons. The bottom line is that
in any version of the  Split SUSY or Little Higgs model 
the effect of cut-off physics is important and as we show in the paper can
produce large corrections to the Higgs mass predictions.

In Ref. \cite{vacsusy} 
the issue of vacuum stability in Split Susy models 
 was addressed by considering the SM vacuum to be not the true vacuum. A constraint on the
 Higgs coupling can then be obtained at the scale $\Lambda$ by requiring that the SM vacuum has not decayed
 to the true vacuum. In our analysis, we consider the SM vacuum to be the true vacuum and require
 the Higgs potential to have a global minimum at the scale $v \sim $ 246 GeV for values of the Higgs field
 $ \stackrel{<}{\sim} \Lambda$. In our analysis we take into account 
the effects of the physics at cut-off in the effective potential which was not considered in Ref. \cite{vacsusy}. The  
issue of vacuum stability in a particular Little Higgs model \cite{Martin}
was also discussed but without the effect of physics at cut-off.

The paper is organized in the following manner:
in Section. 2 we analyze in general terms the requirement of vacuum stability
and the constraints on the Higgs mass. In Section. 3 and Section. 4 we discuss vacuum stability in the 
Split Susy and a specific Little Higgs model, 
and finally in Section. 5 we  summarize our results.

\section{\bf Vacuum Stability- General Analysis} 
We start with a model that is valid up to some cut-off $\Lambda$. Below the 
scale $\Lambda$, are the SM fields as well as additional particles. The 
 physics in this region can be described by an 
   effective theory having the SM gauge symmetry but the 
    lagrangian involves both SM and new particles. The corrections which come from the underlying 
   theory around, and or beyond the cut-off are described by higher dimension 
   operators, 
   \begin{eqnarray} 
   {\cal L}^{new} = \sum_{i} \frac{c_i}{\Lambda^{d_i-4}} {\cal O}^i , \ 
   \label{hd}
   \end{eqnarray} 
   where $d_i$ are the dimensions of ${\cal O}^i$, which are 
   integers greater than 4 and the
 operators ${\cal O}^i$ are 
   $SU(3)_c \times SU(2)_L \times U(1)_Y$ invariant. The dimensionless 
   parameters $c_i$, 
   determining the strength of the contribution of operators ${\cal O}^i$, 
   can be of O(1) or larger. 

    In this work, we are interested in an analysis of the Higgs mass 
   bound from 
   consideration of vacuum stability using the effective lagrangian approach
with higher dimensional operators. 
As discussed in Ref.\cite{DYZ,DZ},
 depending on the size and sign of the coupling of the higher 
   dimensional operator, vacuum stability analysis
 gives a band instead of a single value for
 the lower bound on the Higgs mass for fixed $\Lambda$.
   In
   general, 
   Higgs mass bound from vacuum stability complement the bounds obtained
   from 
   precision electroweak observables. For instance, electroweak precision 
   measurements can be used to obtain an upper bound on the Higgs mass for a
   given 
   $\Lambda$ or alternately for a given Higgs mass one can obtain a lower
   bound on 
   the scale $\Lambda$. Vacuum stability analysis provide a lower bound 
   on the Higgs mass for a given $\Lambda$ and alternately for a given Higgs
   mass 
   one can obtain an upper bound on the scale of new physics $\Lambda$.

   Analyses of the higher dimension 
   operators in Eq. \ref{hd} 
   have been performed by 
   many authors in the literature\cite{bu1,bu2}. 
   The operator, up to dimension 6, 
   relevant for deriving the lower bound on the Higgs mass from vacuum 
   stability is given by 
   \begin{eqnarray} 
   {\cal L}^{new} & = & \frac{c}{\Lambda^2}{( \Phi^+ \Phi - \frac{v^2}{2} )}^3 
   , \ 
\label{hdHiggs}
   \end{eqnarray} 
where the Higgs field $\Phi$ is
\bea
 \Phi &= & \frac{1}{\sqrt{2}}\pmatrix{0  \cr
 \phi} + \frac{1}{\sqrt{2}}\pmatrix{ \sqrt{2} \chi^+ \cr
 i \chi^0}, \
 \label{phidefn}
 \eea
 and $v=<\phi>$ is the vacuum expectation value that minimises the Higgs potential.

The 
   tree level Higgs potential, in the presence 
   of the higher dimensional operator in Eq. \ref{hdHiggs},  can be written as \cite{zhang} 
   \begin{eqnarray} 
   V_{tree} & = & -\frac{m^{2}}{2}\phi^{2} +\frac{1}{4}\lambda \phi^{4} + 
   \frac{1}{8}\frac{c }{\Lambda^{2}} 
   {(\phi^2 -v^2)}^3, \ 
   \label{vtree}
   \end{eqnarray} 
   which is corrected by the one-loop term, $V_{1loop}$, 
   \begin{eqnarray} 
   V_{1loop}(\mu) & = & \sum_{i} \frac{n_i}{64 \pi^{2}}M_i^4(\phi) 
   \left[\log{\frac{M_i^2(\phi)}{\mu^2}} 
   -C_i\right], \
   \label{voneloop} 
   \end{eqnarray} 
   where 
   \begin{eqnarray} 
   M_i^2(\phi) & = & k_i\phi^2 -k_i' \nonumber.\\ 
   \end{eqnarray} 
   The summation goes over the gauge bosons, the 
   fermions and the scalars of the standard model. 
The values of the 
   constants $n_i$, $k_i$, $k_i'$ 
   and $C_i$ can be found in Refs\cite{Casas1,Casas2,Casas3,Casas4,Sher}. 
The full effective potential up to one-loop correction
  is 
   \ber
    V &=& V_{tree} + V_{1loop}.\
    \label{vtotal}
 \eer  

%   Note that the effect of a positive 
%   $c $ is to delay the onset of vacuum instability compared to 
%   the standard model while the effect of a negative 
%   $c $ is to accelerate the onset of vacuum instability. 

   To obtain a lower bound on the Higgs boson mass, 
   in the absence of higher dimensional operators, one 
   can take the location of vacuum instability to be as large as $\Lambda$. 
   However, in our approach, for the low energy theory 
   to make sense
   {\footnote {Effective theory in general will not be valid 
   in the region close to the cutoff scale. One of the examples is the 
   chiral lagrangian of pions where the predictions for processes such as $\pi 
   \pi$ scattering can only be reliable for small momentum transfer relative 
   to the cutoff $\Lambda \sim 4 \pi f_\pi \sim 1$ GeV.}},
    we should require
   $\phi < 
   \Lambda$. We 
   take the scale of vacuum instability, $\Lambda'$, to be $0.5\Lambda$, so 
   the corrections from operators of dimension greater than six to our 
   result is suppressed by a factor of 
   $\frac{{\Lambda'}^2}{{\Lambda}^2} =0.25$. 

   Since we are dealing with values of the field 
   $\phi$ larger than $v$, we need to consider a renormalization group 
   improved potential 
   for our analysis \cite{Casas1,Casas2,Casas3,Casas4, Sher,L,LS,S,Al}. 
As indicated earlier, we have below the scale $\Lambda$, not only the SM fields but additional 
new particles. Hence 
   running for the Higgs self coupling, $\lambda$, the top Yukawa coupling ($g_Y$) will be modified.
Since the effective potential contains one loop correction from the SM fields we use 2-loop
running in the beta functions as far as the SM contributions are involved \cite{Kas1,Kas2}. For the new particles,
their contributions to the beta functions are considered at the one-loop level and the loop effects
of the new particles to the effective potential is neglected. This is not unreasonable as the 
new particles are
higher in mass than the SM particles and the one loop correction from them to the effective potential
will be smaller than the SM contribution.
   The various $\beta$ functions for the SM running to two-loop order 
   can be found in Ref 
   \cite{ME}. 

 In the presence of higher dimensional operators, with the scale of 
 vacuum stability
 $\Lambda'$,
     the vacuum stability requirement
\ber 
   V( \Lambda') & = & V(v) \ 
   \label{vacstab1}
   \eer 
  provides the boundary condition for $\lambda$ at the scale 
   $\Lambda'$, which  using Eq.~\ref{vtotal} and  Eq.~\ref{vacstab1} 
is given by 
   \begin{eqnarray} 
   \lambda_{eff}(\Lambda') & \approx &-\sum_{i} \frac{n_i}{16 
   \pi^{2}}{k_i}^2(\log{k_i}-C_i) -\frac{1}{2}\frac{\Lambda'^2}{\Lambda^2} 
   c + \frac{2m^2}{\Lambda'^2} , 
   \end{eqnarray} 
   where 
   \begin{eqnarray} 
   \lambda_{eff}(\Lambda') & = & \lambda(\Lambda') 
   -\frac{3}{2} c \frac{v^2}{\Lambda^2} . \ 
   \label{vacstab2}
   \end{eqnarray} 
One can then run down the Higgs coupling $\lambda$ to obtain the Higgs mass that ensures
vacuum stability to the scale $\Lambda'$. In general the effective potential 
 increases with $ \phi$ for $\phi >v$ and attains 
a local maximum beyond which it turns around and becomes unstable at the scale $\Lambda'$
where the depth of the potential is the same as for $\phi=v$.

Note that the effect of a positive 
   $c $, in Eq. \ref{hdHiggs}, is to delay the onset of vacuum instability compared to 
   the standard model while the effect of a negative 
   $c $ is to accelerate the onset of vacuum instability. For large enough positive
values of $c$, the effect of the higher dimensional operator can compensate for
the tendency of the standard model Higgs 
potential to become unstable and the instability of the effective potential
disappears for 
 all values of $ v \le \phi \le \Lambda'$. This effect can be demonstrated
in a toy model of new physics  where a scalar field of mass $M$ is added 
to the standard model \cite{Hung}. It was shown in Ref \cite{Hung} that, for 
a given choice of parameters in the effective potential, there is 
a critical value
for the scalar mass ,$M$, below which the vacuum instability disappears.

 In such cases 
the boundary condition
for $\lambda$ at the scale 
 $\Lambda'$, is no longer given by Eq.~\ref{vacstab1} and one 
has to numerically 
search for the
minimum Higgs mass
 that ensures
   \ber 
   V( \phi) & \ge & V(v) \ 
   \eer 
   for all $\phi \le \Lambda'$. 
In fact if one starts from the boundary condition of Eq. (6) the 
vacuum becomes 
unstable much before $\Lambda'$ and the potential attains a second 
local minimum which is 
deeper than the minimum at $\phi=v$\cite{DZ}.   

\section{Vacuum Stability- Split Susy}
In the Split Susy scenario the scale of susy breaking $m_S \equiv \Lambda$
is very high, well beyond a TeV. Below the scale of SUSY breaking,
 are the fermion superpartners, the
Higgsino  ${\tilde H}_{u,d}$,   
the gluino ($\tilde g$), the W-ino ($\tilde W$) 
and B-ino
($\tilde B$), and the SM particles with one Higgs doublet.
Following Ref. \cite{Guidice},
the most general renormalizable
Lagrangian, relevant to our study, with a matter parity, and
gauge-invariant kinetic terms, is given by
\ber
{\cal L}&=&m^2 H^\dagger H-\frac{\lambda}{2}\left( H^\dagger H\right)^2
-\left[ h^u_{ij} {\bar q}_j u_i\epsilon H^* 
+h^d_{ij} {\bar q}_j d_iH
+h^e_{ij} {\bar \ell}_j e_iH \right. \nonumber \\
&&+\frac{M_3}{2} {\tilde g}^A {\tilde g}^A
+\frac{M_2}{2} {\tilde W}^a {\tilde W}^a
+\frac{M_1}{2} {\tilde B} {\tilde B}
+\mu {\tilde H}_u^T\epsilon {\tilde H}_d \nonumber \\
&&\left. +\frac{H^\dagger}{\sqrt{2}}\left( \gtilu \sigma^a {\tilde W}^a
+\gtilup {\tilde B} \right) {\tilde H}_u
+\frac{H^T\epsilon}{\sqrt{2}}\left(
-\gtild \sigma^a {\tilde W}^a
+\gtildp {\tilde B} \right) {\tilde H}_d +{\rm h.c.}\right] ,
\label{lagr}
\eer
where $\epsilon =i\sigma_2$.

The Lagrangian in Eq. \ref{lagr} now should be matched to the
fully supersymmetric  SM  at the scale $\Lambda$ 
which includes the scalars- the
 squarks, sleptons, charged
and pseudoscalar Higgs and is given by
\bea
{\cal L}_{\rm susy}&=&
-\frac{g^2}{8}\left( H_u^\dagger \sigma^a H_u + H_d^\dagger \sigma^a H_d
\right)^2
-\frac{g^{\prime 2}}{8}\left( H_u^\dagger H_u - H_d^\dagger  H_d
\right)^2 \nonumber \\
&&+\lambda^u_{ij}H_u^T\epsilon {\bar u}_i q_j
-\lambda^d_{ij}H_d^T\epsilon {\bar d}_i q_j
-\lambda^e_{ij}H_e^T\epsilon {\bar e}_i \ell_j
\nonumber \\
&&-\frac{H_u^\dagger}{\sqrt{2}}\left( g \sigma^a {\tilde W}^a
+g^\prime {\tilde B} \right) {\tilde H}_u
-\frac{H_d^\dagger}{\sqrt{2}}\left(
g \sigma^a {\tilde W}^a
-g^\prime {\tilde B} \right) {\tilde H}_d +{\rm h.c.}
\label{lagrs}
\eea

The matching allows us to obtain  the coupling constants of the
effective theory
at the scale $\Lambda$.
These are given by \cite{Guidice}
\ber
\lambda(\Lambda )& =& \frac{\left[ g^2(\Lambda )+g^{\prime 2}(\Lambda )
\right]}{4} \cos^22\beta ,
\label{condh}\\
h^u_{ij}(\Lambda ) = \lambda^{u*}_{ij}(\Lambda )\sin\beta , &&
h^{d,e}_{ij}(\Lambda )=\lambda^{d,e*}_{ij}(\Lambda )\cos\beta ,\nonumber\\
\gtilu (\Lambda )= g (\Lambda )\sin\beta ,&&
\gtild (\Lambda )= g (\Lambda )\cos\beta ,\nonumber\\
\gtilup (\Lambda )= g^\prime (\Lambda ) \sin\beta ,&&
\gtildp (\Lambda )= g^\prime (\Lambda )\cos\beta . 
\label{bc}
\eer
The $h_{ij}$'s and the $\lambda_{ij}$'s are the Yukawa couplings in the effective theory
(Eq.~\ref{lagr}) and in the full theory(Eq.~\ref{lagrs}).
The Higgs doublet, $H$ in Eq.~\ref{lagr}  is given by
 $H=-\cos\beta \epsilon H_d^*+\sin\beta H_u$ where
$\tan \beta = v_1/v_2$ with $v_{1,2}$ being the v.e.v
of the neutral component of the Higgs doublet $H_{u(d)}$
satisfying $v=\sqrt{v_1^2+v_2^2} \sim $ 246  GeV.

 As pointed in Ref. \cite{vacsusy} the value of the
 Higgs coupling, $\lambda$, at the scale $\Lambda$ can be easily
modified in the presence of additional terms in the theory at the scale $\Lambda$. In our analysis the boundary condition for the Higgs coupling at 
the scale $\Lambda'$, the scale of vacuum stability, will be provided by the 
requirement of vacuum stability in Eq. \ref{vacstab1}. The boundary
 conditions for the other coupling will be
taken from Eq. \ref{bc}.

\begin{figure}[htb] 
   \centerline{\epsfysize 4.2 truein \epsfbox{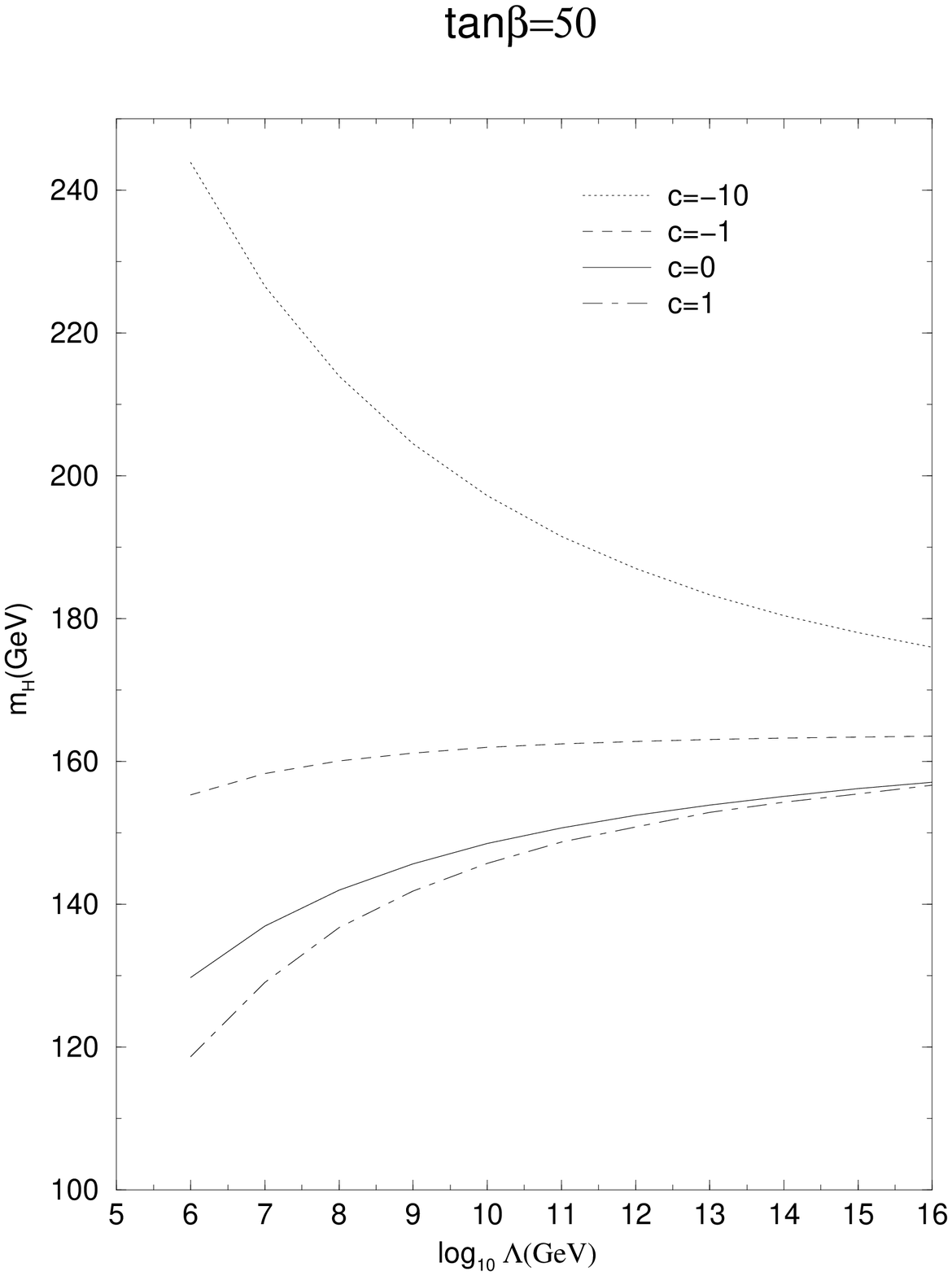}} 
   \caption{The Higgs mass versus the scale of new physics $\Lambda$ for $\tan \beta$=50 in Split Susy} 
\label{fig1}
   \end{figure}

\begin{figure}[htb] 
   \centerline{\epsfysize 4.2 truein \epsfbox{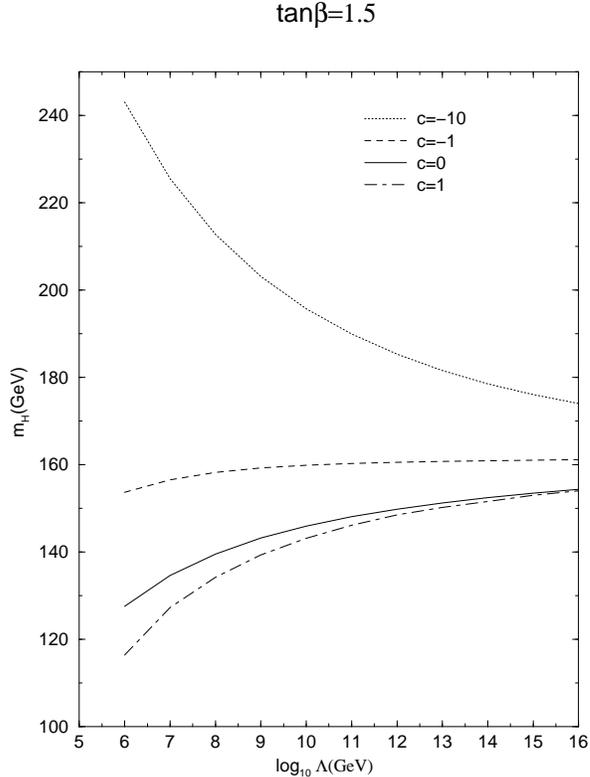}} 
   \caption{The Higgs mass versus the scale of new physics $\Lambda$ for $\tan \beta$=1.5 in Split Susy } 
\label{fig2}
   \end{figure}

In Fig. \ref{fig1} and Fig. \ref{fig2} we plot the Higgs mass versus the scale of SUSY breaking, $\Lambda$.
As we see from the figure the effect of the higher dimensional operator has a significant
effect on the Higgs mass  specially if the sign of $c$ is negative. 
Hence, in the presence of higher dimensional operators, the
Split Susy model can allow a much larger Higgs mass, well beyond the Higgs mass range.
quoted in the literature\cite{Guidice,vacsusy}. For $c=0$ our predictions for the Higgs mass is similar to that obtained
in Ref. \cite{vacsusy}. Notice that the Higgs mass for vacuum stability in this case (c=0)
is bigger than in 
the SM, where a Higgs mass of about 135 GeV ensures vacuum stability up to the Planck scale\cite{Casas1,Casas2,Casas3,Casas4}.
This is because of additional fermionic fields that modify the running of the Higgs coupling.

\section{Vacuum Stability- Little Higgs Model}
One of the attractive features of the Little Higgs models is a natural explanation of electroweak 
symmetry breaking because of the large top Yukawa coupling. Radiative correction involving
a $SU(2)_L$ singlet heavy quark ,T, can provide a negative mass-squared contribution to the
Higgs potential thereby triggering electroweak symmetry breaking.
There are various versions of the Little Higgs model and
we will consider a particular simple version that was proposed in
Ref. \cite{Martin}.
In this simple
 $SU(3)$ model the weak interactions of the Standard Model
are enlarged from $SU(2)\times U(1)$ to $SU(3)\times U(1)$.
There are new gauge bosons associated with the enlarged gauge symmetry
and a new heavy top quark. The quadratic divergences to the Higgs mass from the SM 
gauge bosons and the top quark are canceled by the new 
gauge bosons and the heavy top quark.
 In this model, besides a tree level potential term, the other terms of the Higgs potential are generated 
dynamically through radiative corrections. 
The details of this model are presented in Ref. \cite{Martin} and here we just provide
the bare essentials for the vacuum stability analysis.

The SM gauge group $SU(2)_w \times U(1)_Y$  is enlarged to
$SU(3)_w \times U(1)_X$  and the 
 symmetry breaking,
$SU(3)_w \times U(1)_X \rightarrow SU(2)_w \times U(1)_Y$, occurs when 
 two complex triplet scalar fields
$\Phi_1$ and $\Phi_2$ that transform as $ (1,3)_{-\frac13}$
under $(SU(3)_c,SU(3)_w)_{U(1)_X}$
  get vevs $f_1$ and $f_2$.
After the breaking of $SU(3)_w$  
 the remaining degrees of freedom  in the $\Phi_i$ are conveniently
parametrized in the non-linear representation of the gauge symmetry as   
\bea
\Phi_1= e^{i\Theta  {f_2\over f_1} }
\left( \begin{array}{l}
0  \\ 0 \\ f_1 \end{array} \right) , \quad
\Phi_2= e^{-i \Theta {f_1\over f_2}}
\left( \begin{array}{l}
0  \\ 0 \\ f_2\end{array} \right) 
\label{phiparam}
\eea
where
\bea
\Theta = \frac1{f}\left[{\eta\over \sqrt{2}} + 
\left( \begin{array}{cc} 
\!\!\begin{array}{ll} 0 & 0 \\ 0 & 0 \end{array} 
& \!\!h \\ h^\dagger  & \!\!0 \end{array} \right)\right]
 \quad  {\rm and}\quad  f^2 = f_1^2+f_2^2 \ .
\label{phiexpand}
\eea
The field $h$ is identified 
with the SM Higgs doublet $ \Phi$ defined in Eq.~\ref{phidefn}
and $\eta$ is a real scalar field.

The masses for the gauge bosons, $W_\pm$ and the $SU(2)$ doublet of 
heavy gauge bosons $(W'_\pm, W'_{0})$ are given by 
\ber
m^2_{W_\pm}&=&\frac{g^2}4 v^2 \nonumber \\
m^2_{W'_\pm}&=& m^2_{W'_0}=\frac{g^2}2 f^2, \
\label{gb}
\eer
where $ f^2=f_1^2+f_2^2$.
The neutral gauge bosons have masses
\ber
m_Z^2&\!\!=\!\!& \frac{g^2}4 v^2 (1+t^2)
\nonumber \\
m_{Z'}^2&\!\!\!=\!\!\!& {g^2}f^2 {2\over 3\!-\!t^2} \
\eer
where $t=g'/g=\tan{\theta_W}$ and $\theta_W$ is the weak mixing angle.

There are various ways in which the fermions of the theory can be made to transform
under the enlarged gauge group. 
The masses and mixing of the quarks arise from Yukawa interactions and the form
of the interactions depend on how one chooses to transform the quarks under
$(SU(3)_c,SU(3)_w)_{U(1)_X}$. Within a fairly general scenario that does 
not allow tree level flavour changing neutral current effects one obtains for
 the masses of the up-type quarks and their partners as
\bea
m_u&=&\lambda_u <\!\!h\!\!>
\nonumber \\
m_U&=& \sqrt{(\lambda^u_1 f_1)^2 + (\lambda^u_2 f_2)^2 }
\label{topmassgeneral}
\eea
where  
\bea
\lambda_u&=&\lambda^u_1 \lambda^u_2 \frac{f}{m_{U}} \ ,
\eea 
and $\lambda^u_1$ and $\lambda^u_2$ are Yukawa couplings.
Of interest to us is 
the mass of the heavy top, $m_T$, as it
plays an important role in electroweak symmetry breaking.
The mass of the $m_T$ depends on unknown Yukawa couplings, and so to reduce the number of 
parameters, following Ref. \cite{Martin},  we determine the $\lambda_i^t$
such that the $T$-mass is minimized for given scales $f_i$.
This gives the values
\bea 
\lambda_1 & = & \sqrt{2} \lambda_{t} {f_2 \over f} \nonumber\\
\lambda_2 & = & \sqrt{2} \lambda_{t} {f_1 \over f}, \
\eea
where $\lambda_t$ is the top Yukawa coupling. 
The top and the $T$ quark masses are then given by, 
using Eq.~\ref{topmassgeneral}
\ber
m_t & = & \lambda_t \frac{v}{\sqrt{2}}\nonumber\\
m_T & = & 2 \lambda_{t} {f_1 f_2 \over f},\
\label{topmass} 
\eer
with $v \sim $ 246 GeV is the Higgs vacuum expectations value.

{}For the study of vacuum stability, we do not expect that a more general expression for
$m_T$ in  Eq.~\ref{topmassgeneral}  will significantly alter our results.

The tree level effective potential in this model comes from
\bea
V_{tree}=\mu^2 \Phi_1^\dagger \Phi_2 + h.c. \rightarrow
\mu^2 {f^2\over f_1 f_2} (h^\dagger h+\frac12 \eta^2) 
-\frac1{12} {\mu^2 f^4\over f_1^3 f_2^3} (h^\dagger h)^2 + \dots
\label{treeLH}
\eea
which follows from expanding the $\Phi_{1,2}$ field using 
Eq.~\ref{phiparam} and  Eq.~\ref{phiexpand}.
Including the radiative corrections and replacing $h$ by $h \equiv \Phi$
defined in Eq.~\ref{phidefn}, the effective Higgs potential is given by
\ber
V_{eff} & = & -\frac{1}{2}m^2 \phi^2 + \frac{1}{4}\lambda \phi^4 \nonumber\\
m^2 &=&-(\mu^2 {f^2\over f_1 f_2} + \delta m^2)  \nonumber\\
\lambda & = & ( -\frac{1}{12} {\mu^2 f^4 \over f_1^3 f_2^3} + \delta \lambda) \
\label{Higgspot}
\eer
where
\ber
\delta m^2 &= &\frac{-3}{8\pi^2}\!\!\left[\!\lambda_t^2 m_{T}^2
Log\!\left(\!\frac{\Lambda^2}{m_{T}^2}\!\right)\!
\!-\!\frac{g^2}{4} m_{W'}^2
Log\!\left(\!\frac{\Lambda^2}{m_{W'}^2}\!\right)\!
\!-\!\frac{g^2}{8} (1\!+\!t^2) m_{Z'}^2
Log\!\left(\!\frac{\Lambda^2}{m_{Z'}^2}\!\right)\! \right]\nonumber
\eer
\ber
\delta \lambda  & = & 
\frac{3}{16\pi^2} \left[
\lambda_t^4 Log\!\left(\frac{m_{T}^2}{m_t^2}\right)
-\frac{g^4}{8} Log\!\left(\!\frac{m_{W'}^2}{m_W^2}\!\right)
-\frac{g^4}{16} (1\!+\!t^2)^2 Log\!\left(\frac{m_{Z'}^2}{m_Z^2}\right)
\! \right]  \nonumber \\
& + & {|\,\delta m^2| \over 3} \frac{f^2}{f_1^2 f_2^2} \
\eer

The condition $({dV_{eff} \over d \phi})_v =0$ then gives
\ber
\mu^2 & = & \frac{12f_1^3 f_2^3(\delta m^2+v^2 \delta \lambda)}{f^2(-12 f_1^2 f_2^2+v^2 f^2)}\
\label{mu}
\eer
Now the Higgs effective potential $V_{eff}$ in Eq. \ref{Higgspot} is determined in terms of
the parameters, $f_{1,2}$ and $\Lambda$. The cut-off, $\Lambda$ is $ \sim 4 \pi f_1$ \cite{Martin}
so we will choose $ \Lambda=4 \pi f_1$ to further reduce the parameter space of the model.
Hence for a given value of $\Lambda$, $f_1$ is fixed.
For a given $f_1$ and $m_T$ the Higgs mass is fixed.
For the choice $m_T=1 $ TeV and $f_1=$ 0.5 TeV the Higgs mass is
around 140 GeV.
In our analysis, we will vary $ \Lambda$ between 5-10 TeV. Using Eq. \ref{topmass} we can now solve 
for $f$ in terms of $m_T$,
\ber
 f & = & \frac{2 f_1^2 \lambda_t}{\sqrt{4 f_1^2 \lambda_t^2-m_T^2}}. \
 \label{fmT}
 \eer
We will consider $ f \ge $ 2 TeV to satisfy the electroweak precision measurements\cite{Martin}. As 
$f$ is real, we have the constraint from Eq. \ref{fmT},
 \ber
 m_T & \le & 2\lambda_t f_1. \
 \label{mT}
 \eer
Hence for a given $\Lambda$ and $m_T$, subject to the constraint in Eq. \ref{mT}, the effective potential
 is completely determined
and so is the Higgs mass.

Now we add the effect of physics at or around the cut-off, $\Lambda$,
 via a higher dimensional operator to the Higgs potential in 
Eq. \ref{Higgspot}. The little Higgs model matches on to the SM at the scale
$f$ below which the gauge structure of the model is SM. Above the 
scale $f$  up to the cut off $\Lambda$ the symmetry of the model is
$(SU(3)_c,SU(3)_w)_{U(1)_X}$. The physics at the cut-off can generate
contributions to the effective potential invariant under
$(SU(3)_c,SU(3)_w)_{U(1)_X}$
 of the form
\bea
V_{\Lambda} &= & a_1 M^2( \Phi_1^\dagger \Phi_2) + a_2 ( \Phi_1^\dagger \Phi_2)^2+a_3( \Phi_1^\dagger \Phi_2)^3/\Lambda^2.. + h.c, \
\label{hdlh}
\eea
where $M$ is some mass parameter and $V_{\Lambda}$ is the contribution to the effective potential from
physics at the cut-off $\Lambda$.
 Writing out $V_{\Lambda}$ explicitly in terms of the SM Higgs field $\phi$ we can write
 \bea
V_{\Lambda} &= & A m^2 \phi^2 + B \phi^4+C \phi^6/\Lambda^2. \
\label{hdlh1}
\eea
where m is another mass parameter and the various co-efficients $A,B $ and $C$ are some combination of
$a_{1,2,3}$ and  $f_{1,2}$ with $ \Lambda= 4 \pi f_1$.

Now using the conditions
\bea
({dV_{eff} \over d \phi})_v &=&0 \nonumber\\
({dV_{eff}^2 \over d \phi^2})_v &=&m_H^2, \nonumber\\
\label{renorm}
\eea
allows us to recast $V_{\Lambda}$ in form of the higher dimensional contribution to the effective potential in Eq.~\ref{hdHiggs}.
The full effective potential is then written as,
\begin{eqnarray} 
   V_{eff} & = & -\frac{m^{2}}{2}\phi^{2} +\frac{1}{4}\lambda \phi^{4} + 
   \frac{1}{8}\frac{c }{\Lambda^{2}} 
   {(\phi^2 -v^2)}^3, \ 
\label{vbelowf}
   \end{eqnarray}
The effective potential above should be valid for values of $\phi \le f$.
For a given value of the cut-off $\Lambda$, as we vary $m_T$ in Eq.~\ref{fmT}, 
there is 
a maximum value for $f$ which we call $f_{max}$( Note that $ f \ge 2 $ TeV to 
satisfy the electroweak precision constraints). The value for $f_{max}$ is
 greater than $ \Lambda/2$ for values to $ \Lambda = 5 -10 $ TeV.
Keeping only terms to order dimension  six to represent the effects of cut-off
make sense for values of $\phi$ sufficiently less than $\Lambda$ and so
like the vacuum stability analysis for Split Susy we will choose
$\Lambda'= 0.5\Lambda$ to be the scale of vacuum stability.
Note that $ \Lambda' \sim f_{max}$ and so the form of
Eq.~\ref{vbelowf} to represent the effects of physics around cut-off
is justified.

   The requirement of vacuum stability then gives
   \ber
 \lambda(m_T,\Lambda) & = & 
-\frac{c}{2} \frac{(-\Lambda'^2+v^2)^3}{\Lambda^2 \Lambda'^2(2v^2-\Lambda'^2)} \nonumber \\
& \sim &
-\frac{c}{2} \frac{\Lambda'^2}{\Lambda^2}\,
\label{vaclh}
\eer
where $\Lambda'$ is the scale of vacuum stability discussed above.

As $\lambda(m_T,\Lambda)$ is positive, therefore for positive $c$ the vacuum is always
stable up to the scale $\Lambda'$. For negative values of $c$,
$\lambda(m_T,\Lambda) \ge \frac{|c|}{2} \frac{\Lambda'^2}{\Lambda^2}$ to ensure vacuum stability.
\begin{figure}[htb] 
   \centerline{\epsfysize 4.2 truein \epsfbox{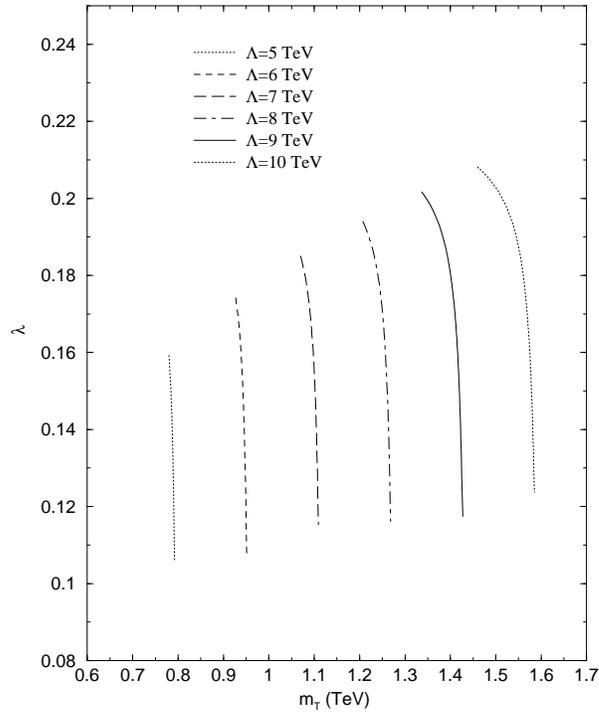}} 
   \caption{The Higgs coupling $\lambda$ versus $m_T$ for various $\Lambda$} 
\label{fig3}
   \end{figure} 
In Fig. \ref{fig3} we plot the Higgs coupling versus $m_T$ for various 
$\Lambda$. 
%This graph shows the range
%of the Higgs coupling, $\lambda$, for various cut off scale $\Lambda$. 
In particular, for $\Lambda=10$ TeV 
the maximum Higgs coupling, $ \lambda_{max }\sim $ 0.21  
which corresponds to a Higgs mass, $m_h=\sqrt{2 \lambda}  v \sim $ 160 GeV. 
 This corresponds to $c \sim 1.7$ from Eq. \ref{vaclh} with 
 the scale of vacuum stability $\Lambda'=0.5\Lambda$. Hence for negative  values of  c, with
  $|c|> 1.7$ the vacuum will become unstable.
For lower values of $\Lambda$ and or $m_T$, smaller $|c|$ will destabilize the vacuum.
Hence, for the Little Higgs model to be viable the corrections from higher dimensional operators have
to be suppressed sufficiently. While we have worked in a specific model our conclusion
is going to be true for different models also, simply due to the fact that the cut-off $\Lambda \sim $ 10 TeV 
or less is 
low and the effect from higher dimensional operators can be quite significant on the effective potential 
and Higgs mass prediction.

\section{Summary}
 In summary, we have studied the effect of higher dimensional operators in the effective potential 
  in two classes of models- the Split Susy models and the Little Higgs models. In both cases the effects of the higher dimensional 
  operators is important when it comes with a negative coefficient. In Split Susy models the Higgs mass 
  predictions
  can be changed significantly specially when the cut-off scale is relatively low. In the Little Higgs models
  where the cut-off scale is much lower, the effect of higher dimensional operator is significant. 
  In a particular Little Higgs model a  value of the coefficient of the higher dimensional operator
  of $O(1)$
  can destabilize the vacuum.

\section{Acknowledgments}

 This work was supported
in part by NSF of China ( X. Zhang)  and by the Natural  Sciences and 
Engineering Research  Council
of Canada (A. Datta ).

   \end{document}